# Uncertainty Factors for Stage-Specific and Cumulative Results of Indirect Measurements

*B. P. Datta*

*Radiochemistry Laboratory, Variable Energy Cyclotron Centre, Kolkata 700 0064, India*

## ABSTRACT

Evaluation of a variable $Y_d$ from certain measured variable(s) $X_i(s)$, by making use of their system-specific-relationship (SSR), is generally referred as the indirect measurement. Naturally, the SSR may stand for a simple data-transformation process ($X_i(s) \rightarrow Y_d$) in a given case, but a set of equations ($f_i(\{Y_d\}) = X_i$, $d$, $i = 1, 2, \ldots, N$), or even a cascade of different such processes ("$X_i(s) \rightarrow {}^{I}Y_d(s) \rightarrow {}^{II}Y_d(s) \ldots \rightarrow Y_d(s)$", or "$X_J \rightarrow {}^{I}Y_d \xrightarrow{X_K} {}^{II}Y_d \ldots \xrightarrow{X_N} Y_d$", or so), in some other case. Further, though the measurements are a priori ensured to be accurate, there is no definite method for examining whether the result obtained at the end of an SSR, specifically a cascade of SSRs, is really representative as the measured $X_i$-*value(s)*.

Of course, it was recently shown that the uncertainty ($\varepsilon_d$) in the estimate ($y_d$) of a specified $Y_d$, viz.: $X_i$*'s* $\rightarrow Y_d$, is given by a specified linear combination of corresponding measurement-uncertainties ($u_i$*'s*). Here, further insight into this principle is provided by its application to the cases represented by cascade-SSRs. It is exemplified how the different stage-wise uncertainties (${}^{I}\varepsilon_d$, ${}^{II}\varepsilon_d \ldots \varepsilon_d$), that is to say the requirements for the evaluation to be successful, could even a priori be predicted. The theoretical tools (SSRs) have resemblance with the real world measuring devices (MDs), and hence are referred as also the data transformation scales (DTSs). However, non-uniform behavior appears to be the feature of the DTSs rather than of the MDs.

## 1. INTRODUCTION

The measurement, which needs to be accomplished through a theoretical process such as the below, is generally labeled as the indirect one:[1]

$$y_d = f_d(x_i) = f_d(X_i + \Delta_i), \quad d = i = 1 \qquad (1a)$$

Or,

$$y_d = f_d(\{x_i\}) = f_d(\{X_i + \Delta_i\}), \quad d = 1 \text{ and } i = 1, 2, \ldots, N = J, K, \ldots, N \qquad (1b)$$

Or, $$f_i(\{y_d\}) = x_i = X_i + \Delta_i, \quad d, i = 1, 2, \ldots, N \qquad (1c)$$

where $y_d$ and $x_i$ stand for the *estimates* of the desired and the actually measured variables, $Y_d$ and $X_i$, respectively, and $\Delta_i$ for the experimental error (the deviation in $x_i$ from its *unknown* true value $X_i$). Clearly, that $Y_d$ is known to so correlate with $X_i(s)$ is the basis for such an evaluation:

$$Y_d = f_d(X_i), \quad d = i = 1 \qquad (1a')$$

Or,

$$Y_d = f_d(\{X_i\}), \quad d = 1 \text{ and } i = 1, 2, \ldots, N = J, K, \ldots, N \qquad (1b')$$

Or,

$$f_i(\{Y_d\}) = X_i, \quad d, i = 1, 2, \ldots, N \qquad (1c')$$

Then any physicochemical evaluation (viz. determination of a characteristic parameter[2,3] or even simply the concentration,[4-6] of a chemical species) should be, it may be pointed out, an example of the indirect measurement only. Further, as indicated above, a given example would be distinguished from another by generally the nature of the system-specific-relationship (SSR) of desired $Y_d(s)$ with the measured $X_i$(s). However, it is possible[6] that one example differs from another only with respect to the desired variable $Y_d$. Thus the evaluations represented, e.g. by the SSRs as "$\boldsymbol{Y_P = X_J / 2}$", "$\boldsymbol{Y_Q = \sqrt{X_J}}$", and "$\boldsymbol{Y_R = (X_J^2 - 1)}$", are different for alone the theoretical tasks described by them are different. However, such an observation makes us enquiring whether all the corresponding results ($y_P$, $y_Q$, and $y_R$) should be equally accurate as the measured estimate ($x_J$). In other words, the question is**:** can the accuracy ($u_i$), to be required in $X_i$–measurements for achieving a preset accuracy ($\varepsilon_d$) in the desired result ($y_d$), be dictated by the SSR shaping $y_d$?

However, it may be reminded, the evaluation of error in an unknown-estimate (either $x_i$, or $y_d$) is impossible. Again, there could be no alternative to the assessment of the possible error for ensuring whether the corresponding measurement is useful. Thus, for any measurement ($X_i$) to be carried out, it is a well-known norm to first develop the method to the extent that the maximum possible value (MPV) of experimental error ($\Delta_i$) should at least be acceptable. The MPV of error is referred to here as, it should be noted, either[1] uncertainty or[7] accuracy (inaccuracy). Further, we consider the error as relative only, i.e.: $\Delta_i = (\Delta X_i / X_i) = ([x_i - X_i] / X_i)$, and then: $u_i = |{}^{Max}\Delta_i|$. Similarly, denoting the error in the result $y_d$ as $\delta_d$, we mean that: $\delta_d = (\delta Y_d / Y_d) = ([y_d - Y_d] / Y_d)$, and: $\varepsilon_d = |{}^{Max}\delta_d|$. Now, returning to above, it may be mentioned that the point at issue has already been evaluated[7]. It was thus clarified that the shaping of a result $y_d$ should also mean the fixation of its uncertainty $\varepsilon_d$ **by the SSR** involved. That is to say that, as any oxidation reaction but without the complementary reduction process is inconceivable, the desired (i.e. a given SSR dictated **systematic**) change as "$x_i(s) \rightarrow y_d$" but without really the complementary uncertainty(s) transformation "$u_i(s) \rightarrow \varepsilon_d$" is also unthinkable. In short, it has previously been explained[7] why the uncertainty $\varepsilon_d$ can even numerically be different from the uncertainty $u_i$, i.e. why the SSR can have a say in planning the required experiments. At present, the idea is simply to elaborate on the implications of such facts for cases of somewhat *involved* SSRs.

Actually, the result shaping: $x_i(s) \rightarrow y_d$ is in many a case represented by (instead of Eq. 1a, or Eq. 1b, or Eq. 1c, alone) a cascade of computational processes (COCP) comprising one or more types as Eqs. 1a-1c. Moreover, a study may require different kinds of experiments (variables: $X_J$, $X_K$, … $X_N$), and hence the different estimates ($x_J$, $x_K$, … $x_N$) would likely be subject to different uncertainties ($u_J$, $u_K$, …$u_N$). Even, different $x_i$'s might conform the inputs to different stages of a COCP (viz.: $X_J \rightarrow {}^{I}Y_d \xrightarrow{X_K} {}^{II}Y_d \ \ldots \ \xrightarrow{X_N} Y_d$ ). Further, as pointed out above,[7] any output (e.g. ${}^{I}y_d$, but which should be an **input** to the 2nd stage of the COCP) will even for purely random sources of errors in the estimates ($x_i$*'s*) of $X_i$*'s* be subject to**:** (i) systematic-error, and (ii) to the extent decided by the corresponding (1st stage) SSR. That is, assessing (predicting) uncertainty in a result obtained via a COCP appeals difficult, but is defined the objective here.

In fact, an evaluation as Eq. 1 might be (for computational convenience and/ or for specific objectives) designed as a COCP. For example, in the case of determining a light element isotopic ratio ($Y_d$) by mass spectrometry (IRMS), the**:** $X_i(s) \rightarrow Y_d(s)$ transformation is with the idea to avoid the reporting of results on **non-corresponding scales**[8] carried out via a COCP. That is, as recommended,[8,9] the input(s) for and the output(s) from an Eq. 1 representing the IRMS system are made to be subject to certain scale-conversions. However, it is generally difficult to ascertain whether the scale-conversion processes used there in IRMS really assure the purpose. However, the uncertainty-consideration here is believed to help crosscheck the pros and cons for involving any data-translation process (an Eq. 1 or a COCP) in a basically experimental study.

The work is organized as follows. First, we make a simplifying consideration of terminologies cum principles (section 2). Subsequently (section 3), we discuss our findings.

Here, it may be mentioned that a given SSR (an Eq. 1) is sometimes for convenience referred to by alone $Y_d$, viz. "$Y_P = X_J / 2$" by "$Y_P$".

## 2. PRINCIPLES

### 2.1 Data Transformation Scale (DTS)

It may be pointed out that, by the behavior, an SSR is indistinguishable from a measuring device (MD). For example, as the response of an MD depends on the quantity being measured, $Y_d$ (or estimate**:** $y_d$) will always vary as a function of $X_i$ (or estimate**:** $x_i$, cf. Eq. 1). Moreover, as the response of an MD towards a given quantity is decided by its own feature, the output ($Y_d$) for any given input ($X_i(s)$) here will depend on the SSR itself, e.g. $X_J$ gets projected as however three different outputs by the above SSRs: $Y_P = X_J/2$, $Y_Q = \sqrt{X_J}$, and $Y_R = (X_J^2 - 1)$, which are also therefore referred to as the three different data transformation scales (DTSs).

***2.1.1 Characteristics of a DTS: non-uniform behavior***

A DTS can like any usual MD be shown bracketed with certain parameter(s) $M_i^d(s)$ dictating its behavior,[7] e.g. an Eq. 1b/ could be marked with:

$$M_i^d = \left(\frac{\partial Y_d}{\partial X_i}\right)\left(\frac{X_i}{Y_d}\right) = \left(\frac{\partial Y_d/Y_d}{\partial X_i/X_i}\right), \quad d = 1, \text{ and } i = 1, 2, \ldots N \text{ (or, } i = J, K, \ldots N) \qquad (2)$$

Similarly an Eq. 1c/, which consists of $N$ different SSRs (with: $d$, $i$ = 1, 2 **…** $N$), will naturally have so many parameters ($M_i^d$ *'s*) as $N^2$. Of course, "$N$" is unity for cases represented by Eq. 1a/. Yet, it should generally be possible to distinguish a given Eq. 1a/, in terms the characteristic **rate** ($M_i^d$) of relative variation in $Y_d$ as a function of $X_i$, from another. For example, $Y_P$, $Y_Q$ and $Y_R$ above are predicted to differently vary (with $X_J$): $M_J^P$ = 1.0 (cf. $Y_P$); $M_J^Q$ = ½ (cf. $Y_Q$); and $M_J^R = 2X_J^2/(X_J^2 - 1)$. That is to say that $Y_P$ will (for a given change in $X_J$) vary at a rate different from that of $Y_Q$, but the rate is in either case ever **fixed**. However, the response of $Y_R$ will be decided by $X_J$ itself. For illustration, let: $X_J$ **= 0.1**. Then, one can verify that: $M_J^R$ = -0.0202 (cf. Eq. 2). Now, say: $X_J$ **= 1.1**, which however yields: $M_J^R$ = 11.5238.

Further, it may be mentioned that uniformity-in-response stands generally as a useful criterion for recognizing an MD. In addition, the results of measuring a given quantity by two appropriate but different kinds of MDs should by and large be expected the same. However, the behavior of even alone the DTS as $Y_R$ (viz. $M_J^R$ depends on $X_J$) should suffice explaining why *non-uniformity* as a feature be better attributed to the DTS than to the MDs.

### *2.1.2 Uncertainty transfer via a DTS (an Eq. 1)*

The signature for a relationship (Eq. 1) between $X_i(s)$ and $Y_d(s)$ is that the substitution for $X_i(s)$ by any desired ${}^{True}x_i(s)$ should yield the corresponding ${}^{True}y_d(s)$, and the vice versa. However, this had led to the query[7] whether $x_i(s)$ having subject to say 0.05% uncertainty will cause the desired $y_d(s)$ to be at exactly 0.05% uncertainty (i.e. should $\varepsilon_d$ equal $u_i$?), which was in turn evaluated by introducing the above term "behavior of a DTS ($M_i^d$, cf. Eq. 2)". Thus, it was clarified that:[7]

$$\varepsilon_d = \sum_{i=J}^{N} \left|M_i^d\right| u_i \qquad (3)$$

Or, if all inputs ($x_i$'s) are subject to a given uncertainty (${}^{G}u$), i.e. if $u_i$ = ${}^{G}u$ ($i$ = $J, K, \ldots N$), then:

$$\varepsilon_d = (\sum_{i=J}^{N} \left|M_i^d\right|)\, u_i = (\sum_{i=J}^{N} \left|M_i^d\right|)\, {}^{G}u \qquad (4)$$

Here, it may also be pointed out that Eq. 2 defines $M_i^d(s)$ as the theoretical constant(s) for a given DTS, thereby enabling even a priori prediction of the uncertainty ($\varepsilon_d$). Further, for a simple case as Eq. 1a$'$ ($d = i = N = 1$), Eq. 3/ 4 reduces to:

$$\varepsilon_d = \left|M_i^d\right| u_i = \left|M_i^d\right| {}^{G}u \qquad (3a)$$

It may be noted that neither the evaluation of the uncertainty $\varepsilon_d$ requires the knowledge of the $u_i$-sources, nor can $\varepsilon_d$ vary depending on whether $u_i$ stands for random and/ or systematic causes. Further, Eq. 3 should stand exact for all cases represented by linear SSRs only.[7] However, as the factors of "$(u_i)^P$, with: $P \geq 2$" or so are ignored[7] in Eq. 3, the uncertainty ($\varepsilon_d$) corresponding to a non-linear DTS (and for finite $u_i(s)$) might in reality vary from that given by Eq. 3. Nevertheless, the experiments are always so designed that: $u_i \rightarrow$ zero. Therefore, Eq. 3 should also suffice explaining the non-linear cases. In support, we offer the evaluation below.

Let the test-systems to be represented by the SSRs $Y_P$, $Y_Q$ and $Y_R$ above, and an $X_J$-standard by "${}^{T}X_J$ = ${}^{True}x_J$ = 0.1 (i.e. say: $Y_P$ = 0.05; $Y_Q$ = 0.3162278; and: $Y_R$ = -0.99)". Further, let the method of measurement be so established that: $u_J$ = 0.05% (i.e. say the estimates of ${}^{T}X_J$ were observed to be so restricted as: $0.10005 \geq {}^{T}x_J \geq 0.09995$). Then, on the one, Eq. 3a predicts: $\varepsilon_P = |M_J^P| u_J = u_J = \mathbf{0.05\%}$ (as: $M_J^P = 1$, cf. section 2.1.1); $\varepsilon_Q = |M_J^Q| u_J = u_J/2 = \mathbf{0.025\%}$; and $\varepsilon_R = |M_J^R| u_J = 0.0202u_J = \mathbf{0.00101\%}$. On the other, taking e.g. the *lowest* estimate: ${}^{T}x_J = 0.09995$, one obtains: $y_P = 0.049975$, $y_Q = 0.316149$, and: $y_R = -0.99001$. Thus, it may be noted that not alone "$|\delta_P| = |{}^{Max}\delta_P| = |(y_P - Y_P)/Y_P| = \mathbf{0.05\% = \varepsilon_P}$"; but "$|{}^{Max}\delta_Q| = \mathbf{0.025003\% \cong \varepsilon_Q}$"; or "$|{}^{Max}\delta_R| = \mathbf{0.00101\% = \varepsilon_R}$". Similarly, let the measurement-technique be so independent of $X_J$ that, against "${}^{T}X_J = 1.1$", it yields: $1.10055 \geq {}^{T}x_J \geq 1.09945$ (i.e. again suppose that: $u_J = 0.05\%$). Then, taking e.g. the *highest* estimate (${}^{T}x_J = 1.10055$), one obtains: $y_P = 0.550275$ (i.e.: $|{}^{Max}\delta_P| = \mathbf{0.05\% = \varepsilon_P}$); $y_Q = 1.049071$ (i.e.: $|{}^{Max}\delta_Q| = \mathbf{0.024997\% \cong \varepsilon_Q}$); and: $y_R = 0.21121$ (i.e.: $|{}^{Max}\delta_R| = \mathbf{0.5763\% \cong \varepsilon_R}$), which further verify the theory that (though: $\varepsilon_P = \boldsymbol{u_J}$, or: $\varepsilon_Q = \boldsymbol{u_J/2}$) $M_J^R$, and hence $\varepsilon_R$, will *vary* with $X_J$ ($\varepsilon_R = 11.5238u_J = \mathbf{0.5762\%}$, cf. Eq. 3a).

## 2.2 $\mathbf{u_i(s) \rightarrow \varepsilon_d}$ via a COCP: stage-wise uncertainty factors (UFs)

Suppose that the parameter ${}^{III}Y_d$ of an arbitrary system is determined via a COCP as:

$$ {}^{I}Y_d = f_d(X_J, X_K) = X_J - X_K \qquad (5) $$

$$ {}^{II}Y_d = f_d({}^{I}Y_d) = \alpha\,({}^{I}Y_d + 1) \qquad (5a) $$

$$ {}^{III}Y_d = f_d({}^{II}Y_d, X_L) = {}^{II}Y_d + X_L \qquad (5b) $$

where the prefix as "I" refers to the stage-number, and α is a constant. Clearly, the desired variable (${}^{III}Y_d$) could be obtained as a single-stage-process. However, the idea is to also represent the cases, where: (i) not alone the COCP output (here, ${}^{III}Y_d$) but even an inter-stage one (${}^{I}Y_d$ or ${}^{II}Y_d$) is a desired variable,[2,3] (ii) a single-step process is, say simply for convenience, split up into more than one,[10,11] etc. Anyway, the basic purpose here is to examine whether the inter-stage *theoretical-processes* can cause their output-parameters as ${}^{I}\varepsilon_d$, ${}^{II}\varepsilon_d$, and ${}^{III}\varepsilon_d$, to numerically differ from one another, and/ or whether should it be a fact that**:** $u_i < {}^{I}\varepsilon_d < {}^{II}\varepsilon_d < {}^{III}\varepsilon_d$, or so.

The 1st stage process, it may be pointed out, can in no way be different from an individual DTS, and hence the uncertainty (${}^{I}\varepsilon_d$) in the result (${}^{I}y_d$) should be obtained as (cf. Eq. 3):

$$ {}^{I}\varepsilon_d = \left|{}^{I}M_J^d\right| u_J + \left|{}^{I}M_K^d\right| u_K = \left(\left|{}^{I}M_J^d\right| F_J + \left|{}^{I}M_K^d\right| F_K\right) {}^{G}u = {}^{I}[UF]_d\, {}^{G}u \qquad (6) $$

where**:** ${}^{I}M_J^d$ and ${}^{I}M_K^d$ stand for the 1st stage SSR-specific parameters (cf. Eq. 2), $u_i$*'s* for the measurement-uncertainties, ${}^{G}u$ for any preset value (e.g. 0.01%) of "$u_i$" so that**:** $F_J = (u_J / {}^{G}u)$, $F_K = (u_K / {}^{G}u)$, and/ or the 1st stage uncertainty factor (${}^{I}[UF]_d$) is a priori predicted as:

$$ {}^{I}[UF]_d = \Sigma^{K}{}_{i=J} \left|{}^{I}M_i^d\right| F_i = \left|{}^{I}M_J^d\right| F_J + \left|{}^{I}M_K^d\right| F_K \qquad (7) $$

The 2nd stage has only one input (${}^{I}y_d$) but is generated by the 1st stage. Otherwise, the 2nd-stage (i.e. the 1st cumulative) uncertainty ${}^{II}\varepsilon_d$ should also be obtained as Eq. 3/ 3a:

$$ {}^{II}\varepsilon_d = \left|{}^{II}M_d^d\right|\ {}^{I}\varepsilon_d = \left|{}^{II}M_d^d\right|\ {}^{I}[UF]_d\ {}^{G}u = {}^{II}[UF]_d\ {}^{G}u \qquad (6a) $$

where**:** ${}^{II}M_d^d$ is the characteristic theoretical constant of the 2nd stage SSR, and ${}^{II}[UF]_d$ is called as the 1st cumulative (or, 2nd stage) uncertainty-factor and is also a theoretical parameter:

$$ {}^{II}[UF]_d = \left|{}^{II}M_d^d\right| {}^{I}[UF]_d = \left|{}^{II}M_d^d\right|\ (\Sigma^{K}{}_{i=J} \left|{}^{I}M_i^d\right| F_i) \qquad (7a) $$

The 3rd stage (Eq. 5b) is analogous to the 1st stage, but its inputs comprise the inter-stage-dependent variable ${}^{II}Y_d$, and an independent one ($X_L$). Hence, the output-uncertainty (${}^{III}\varepsilon_d$) should here again be obtained as a linear combination of input uncertainties (${}^{II}\varepsilon_d$ and $u_L$):

$$ {}^{III}\varepsilon_d = \left|{}^{III}M_d^d\right|\ {}^{II}\varepsilon_d + \left|{}^{III}M_L^d\right|\ u_L = \left|{}^{III}M_d^d\right|\ {}^{II}[UF]_d\ {}^{G}u + \left|{}^{III}M_L^d\right|\ F_L\ {}^{G}u = {}^{III}[UF]_d\ {}^{G}u \qquad (6b) $$

where**:** ${}^{III}M_d^d$ and ${}^{III}M_L^d$ stand for the predicted rates of variations (cf. Eq. 2) of ${}^{III}Y_d$ as a function of ${}^{II}Y_d$ and $X_L$, respectively, $F_L = (u_L / {}^{G}u)$, and ${}^{III}[UF]_d$ is the 2nd cumulative uncertainty-factor:

$$^{III}[UF]_d = \left|{}^{III}M_d^d\right| {}^{II}[UF]_d + \left|{}^{III}M_L^d\right| F_L \quad (7b)$$

Thus, as shown, Eq. 3 stands as the fundamental expression of uncertainty ($\varepsilon_d$) in the result $y_d$. It may however be mentioned that, only for indicating the process involved is a COCP, the final output is also prefixed here (cf. $^{III}Y_d$). Further, it may be noted that $\varepsilon_d$ can, depending upon the value(s) of the SSR-specific constant(s) as $M_i^d(s)$ and/ or the COCP-parameters as $[UF]_d$, turn out $>{}^{G}\mathbf{u}$ (or, $>\boldsymbol{u_i}$), equal to $^{G}\mathbf{u}$, or even $<{}^{G}\mathbf{u}$,. Now, let us imagine that: $u_J = u_K = u_L = {}^{G}u$, so that the factors as $F_J$, $F_K$ and $F_L$ be all unity, and/ or that the uncertainty-factors (*UFs*) be decided by the stage-related $M_i^d$*'s* only(cf. Eq. 7c below). Clearly, even then, no trend as either: $^{III}\varepsilon_d = {}^{II}\varepsilon_d = {}^{I}\varepsilon_d = {}^{G}u$, or: $^{G}u < {}^{I}\varepsilon_d < {}^{II}\varepsilon_d < {}^{III}\varepsilon_d$, or so, should stand for a general fact.

$$\begin{aligned} ^{III}[UF]_d &= \left|{}^{III}M_d^d\right| {}^{II}[UF]_d + \left|{}^{III}M_L^d\right| F_L = \left|{}^{III}M_d^d\right| \left|{}^{II}M_d^d\right| {}^{I}[UF]_d + \left|{}^{III}M_L^d\right| F_L \\ &= \left|{}^{III}M_d^d\right| \left|{}^{II}M_d^d\right| (\left|{}^{I}M_J^d\right| F_J + \left|{}^{I}M_K^d\right| F_K) + \left|{}^{III}M_L^d\right| F_L \\ &= \left|{}^{III}M_d^d\right| \left|{}^{II}M_d^d\right| (\left|{}^{I}M_J^d\right| + \left|{}^{I}M_K^d\right|) + \left|{}^{III}M_L^d\right| \end{aligned} \quad (7c)$$

## 3. COCP SYSTEMS: VERIFICATION OF THE UNCERTAINTY THEORY

### 3.1 The arbitrary COCP system as Eqs. 5-5b

Whether our treatment above is correct or not can be ascertained provided the system-specific $X_i$*-values*, and hence the $^{I}Y_d$, $^{II}Y_d$, etc., are known. Therefore, instead of a real world system (with unknown $X_i$*'s*), we first consider the COCP system as Eqs. 5-5b to be represented by, say: $\alpha = 0.25$; and by the $X_J$, $X_K$ and $X_L$-standards as: $^{T}X_J = 5.0$, $^{T}X_K = 3.0$ and $^{T}X_L = 0.50$, respectively. We also presume that all measurements were desired to be 0.05% accurate (i.e. $^{G}u = 0.05\%$).

However say, while the measurements of $X_J$ and $X_K$ had actually ensured: $u_J = u_K = 0.05\%$ (i.e. tough the repetitive measurements of $^{T}X_J$ and $^{T}X_K$ had yielded: $4.9975 \geq {}^{T}x_J \geq 5.0025$, and: $2.9985 \geq {}^{T}x_K \geq 3.0015$, respectively), the $X_L$-measurement was tricky as that: $u_L = 0.4\%$ (i.e.:

$0.5002 \geq {}^{T}x_L \geq 0.498$). Then, it should be noted that: $F_L = (u_L / {}^{G}u) = 8$, but: $F_J = F_K = 1$. However the stage-specific parameters ($M_i^d(s)$, $[UF]_d$, and $\varepsilon_d$) are presented in Table 1 (cf. **Block No. 1**), which clarifies that $M_i^d$*'s* (e.g. ${}^{I}M_J^d$, and ${}^{I}M_K^d$), and hence $[UF]_d$*'s* and/ or $\varepsilon_d$*'s*, depend on $X_i$*'s*. That is to say that, if $X_i$*'s* vary from their chosen true values (${}^{T}X_i$*'s*) above, then ${}^{I}\varepsilon_d$, ${}^{II}\varepsilon_d$ and ${}^{III}\varepsilon_d$ would also vary from their values predicted in Table 1.

However, how should we verify, e.g. that the 1st stage output-uncertainty (${}^{I}\varepsilon_d$) is in reality 0.2%? Clearly, the same is true, provided[7]: (i) only **two** specific combinations of the (highest and the lowest) estimates of ${}^{T}X_J$ and ${}^{T}X_K$ above should cause the result (${}^{I}y_d$) to be at ±0.2% error; and (ii) **all other** (${}^{T}x_J$ and ${}^{T}x_K$) combinations should imply "$|{}^{I}\delta_d| < 0.2\%$". Nevertheless, it could be shown that: (1) ($x_J = {}^{T}x_J = 5.0025$ and $x_K = {}^{T}x_K = 2.9985$) give "${}^{I}y_d = 2.004$, i.e.: ${}^{I}\delta_d$ = **0.2%**"; and (2) ($x_J = 4.9975$ and $x_K = 3.0015$) yield "${}^{I}y_d = 1.996$, i.e.: ${}^{I}\delta_d$ = **-0.2%**"; but even (3) ($x_J = 5.0025$ and $x_K = 3.0015$) give an output equally accurate as them (${}^{I}y_d = 2.001$, i.e.: ${}^{I}\delta_d$ = **0.05%** = $|\Delta_i|$). Further: (4) $x_J = 5.001$ (with: $\Delta_J = 0.02\%$) and $x_K = 3.001$ (with: $\Delta_K = 0.0333\%$), yield "${}^{I}y_d$ = **2.0** = ${}^{I}Y_A$". This, it may be noted, clarifies the fact[7] that "**$\Delta_i$'s ≠ 0.0**" can cause "**$\delta_d$ = 0.0**". Actually, the corresponding predicted requirement[7] is: $\Delta_J/\Delta_K = -({}^{I}M_K^d / {}^{I}M_J^d) = 0.6$ (cf. Table 1), which is satisfied in the present case (as: $\Delta_J/\Delta_K = 0.6$). Over and above, the uncertainty factor has also the value as predicted: ${}^{I}[UF]_d = ({}^{I}\varepsilon_d / {}^{G}u) = (0.2 / 0.05) = 4.0$ (cf. Table 1).

Now, coming to the 2nd stage process (Eq. 5a), it could be readily seen that: **(i)** ${}^{I}y_d = 2.004$ (cf. example no. **1** above) yields "${}^{II}y_d = 0.751$ (i.e.: ${}^{II}\delta_d = 0.133\%$)"; **(ii)** ${}^{I}y_d = 1.996$ gives "${}^{II}y_d = 0.749$ (with: ${}^{II}\delta_d = -0.133\%$)"; but **(iii)** the example nos. **3** and **4** give "${}^{II}y_d = 0.75025$", and "${}^{II}y_d = 0.75$", respectively (i.e. $|{}^{II}\delta_d| < 0.133\%$). These verify that: ${}^{II}\varepsilon_d = 0.133\%$, and/ or: ${}^{II}[UF]_d = {}^{II}\varepsilon_d/{}^{G}u =$ 2.667 (cf. Table 1). Even this prediction (${}^{II}\varepsilon_d$ = **0.133%**) could be, as indicated above, shown true for the stages I and II together to be the single DTS**:** ${}^{II}Y_d = f^{II}(X_J, X_K) = \alpha\ (X_J - X_K + 1)$.

Similarly it can be supplemented that the output ${}^{III}y_d$ will turn out more **inaccurate** than either the 1st (${}^{I}y_d$) or the 2nd (${}^{II}y_d$) stage-output (${}^{III}\varepsilon_d$ = **0.24%**, cf. Table1 ), viz. only the two pairs of input-estimates as (${}^{II}y_d = 0.751$ and $x_L = 0.502$) and (${}^{II}y_d = 0.749$ and $x_L = 0.498$) yield**:** ${}^{III}y_d$ = (${}^{III}Y_d \pm$ **0.0024** ${}^{III}Y_d$). However, all other possible pairs of ${}^{II}y_d$ and $x_L$ imply**:** $|{}^{III}\delta_d| < 0.24\%$ (e.g. ${}^{II}y_d = 0.75025$ and $x_L = 0.502$, gives**:** ${}^{III}y_d = 1.25225$ i.e. ${}^{III}\delta_d = 0.18\%$). Here again, the facts (${}^{III}\varepsilon_d$ = 0.24%, and/ or**:** ${}^{III}[UF]_d = {}^{III}\varepsilon_d / {}^{G}u = 4.8$) could be shown to remain unchanged for considering**:** ${}^{III}Y_d = f^{III}(X_J, X_K, X_L) = (\alpha\ (X_J - X_K + 1) + X_L)$.

However, if**:** $F_J = F_K = F_L = 1$ (i.e. if also**:** $\boldsymbol{u_L = {}^{G}u}$), then Eq. 7c predicts**:** ${}^{III}[UF]_d = 2.0$, which in turn gives**:** ${}^{III}\varepsilon_d = 2\,{}^{G}u = 0.1\%$. That is to say that the estimate ${}^{III}y_d$ would **then** be more **accurate** than either the estimate ${}^{I}y_d$ or ${}^{II}y_d$ (as, it may be pointed out, Eq. 5 or Eq. 5a does not involve $X_L$, i.e. as $F_L$ cannot affect ${}^{I}\varepsilon_d$ (**0.2%)** and ${}^{II}\varepsilon_d$ (**0.133%**)). This further clarifies the basic fact that any output uncertainty $\varepsilon_d$ (and/ or *[UF]*) would jointly be decided by the corresponding SSR and its input-uncertainties ($u_i(s)$ and/ or $\varepsilon_d$(s), i.e. as the case may be). In addition, the finding such as that neither**:** ${}^{I}\varepsilon_d = u_i$ (= ${}^{G}u$ = 0.05%), nor**:** ${}^{I}\varepsilon_d = {}^{II}\varepsilon_d = {}^{III}\varepsilon_d$, is a confirmation that even a linear DTS may cause its output to vary by accuracy from its input.

### 3.2 Standard free energy ($Y_G$) of micellization

The free energy of mecellization $Y_G$ was evaluated[2] via the critical concentration ($Y_C$) as:

$$Y_C = f_C(X_J, X_K, X_L) = X_J / (X_K - X_L) \tag{8}$$

and,

$$Y_G = f_G(Y_C) = R\ T\ ln(V_0\, Y_C\ /\ [1 + V_0\, Y_C\ ]) \tag{8a}$$

where $X_i$*'s* stand for relevant experimental variables, $R$ for the gas constant, $T$ for absolute temperature, and $V_0$ is a constant of reaction medium.

However we enquire whether $Y_G$, which was obtained as a function of $Y_C$ alone, is exactly as accurate as the estimate ($y_C$) of $Y_C$. We therefore work out the parameters of the COCP-system here, and present them in Table 1 (cf. **BLOCK No. 2**). The table clarifies that the behavior (i.e.**:** $M_i^d(s)$) of either the 1st or the 2nd stage process would depend on $X_i$*'s* but which are unknown. However, for the specific experimental conditions referred[2] to as "$R_w$ = [water]/[surfactant] = 1" in chloroform ($V_0$ = 0802 $dm^3\ mol^{-1}$), the $X_i$*-values* obtained**:** $x_J = 0.22 \pm 0.04$, $x_K = 18.4 \pm 0.1$ (in $dm^3\ mol^{-1}$) and $x_L = 10.6 \pm 0.6$ (in $dm^3\ mol^{-1}$). These give**:** $y_C$ = 28.20 (reported[2]**:** 28.3) mmol $dm^{-3}$, which in turn yields**:** $y_G$ = -15.1 kJ $mol^{-1}$. It should be noted that such a set of the measured estimates ($x_i$*'s*) are taken here as the ${}^{true}x_i$*'s* (i.e.**:** $X_i = x_i$, and hence**:** $Y_d = y_d$), and their scatters as the measurement-uncertainties (i.e.**:** $u_J = [(0.04 / x_J) = (0.04 / 0.22)] = 18.2\%$, $u_K = 0.54\%$, and $u_L = 5.7\%$), and accordingly the behavior of the system is elaborated in Table 1. Further, as $u_i$*'s* are known, the consideration and non-consideration of the parameters as $F_i$*'s* (cf. section 2.2) should make no difference. Yet, in Table 1 (Block no. 2), the uncertainty-evaluation is illustrated for the preset measurement-uncertainty (${}^{G}u$) of 0.5% (i.e. $\boldsymbol{{}^{G}u \approx u_K}$), and hence for**:** $F_J = (u_J / {}^{G}u) = 36.4$, $F_K = 1.08$, and $F_L = 11.4$.

Table 1 predicts the *estimates* of $Y_C$ and $Y_G$ to be subject to the uncertainties as high as **≈50** and **≈9** times the least experimental-uncertainty $u_K$, respectively. Thus, for example, it could be shown that "$x_J = (X_J + 18.2\%)$, $x_K = (X_K - 0.54\%)$, and $x_L = (X_L + 5.7\%)$" yield "$y_C = (Y_C + 30\%)$, i.e. $|{}^{Max}\delta_C| = 30\%$, and hence $[UF]_C = (|{}^{Max}\delta_C| / {}^{G}u) = 60$"; which in turn yields**:** $y_G = (Y_G - 4.28\%)$, i.e. $[UF]_G = (4.28 / 0.5) = 8.56$. Clearly, the reason is that the $u_i$*'s* are generally high. Nevertheless, that the predictions (Table 1) are sound could be better verified by scaling down the errors by say a factor of 100 (i.e. say**:** ${}^{G}u = 0.005\%$, so that**:** $\varepsilon_C$ = **0.272%** and**:** $\varepsilon_G$ = **0.0445%**), viz.**: (1)** $x_J = (X_J + 0.182\%)$, $x_K = (X_K - 0.0054\%)$, and $x_L = (X_L + 0.057\%)$, yields "$y_C$

= ($Y_C$ **+ 0.27%**), i.e. $[UF]_C$ = (0.27 / $^Gu$) = 54", which in turn gives "$y_G$ = ($Y_G$ **– 0.045%**), i.e. $[UF]_G$ = 9"; and **(2)** $x_J$ = ($X_J$ - 0.182%), $x_K$ = ($X_K$ + 0.0054%), and $x_L$ = ($X_L$ - 0.057%), imply "$y_C$ = ($Y_C$ **- 0.27%**), i.e. $[UF]_C$ = 54", and "$y_G$ = ($Y_G$ **+ 0.045%**), with $[UF]_G$ = 9". Further, all other error-combinations with "$|\Delta_J| \le 0.182\%$, $|\Delta_K| \le 0.0054\%$, and $|\Delta_L| \le 0.057\%$" could be shown to cause**:** $|\delta_C|$ **< 0.27%** and, $|\delta_G|$ **< 0.045%**, e.g. **(3)** $x_J$ = ($X_J$ + 0.182%), $x_K$ = ($X_K$ - 0.0054%), and $x_L$ = ($X_L$ - 0.057%), yield "$y_C$ = ($Y_C$ + 0.12%), and "$y_G$ = ($Y_G$ - 0.02%).

However, in order for having a better picture as to how $u_i$*'s* affect the desired results, we now consider all the three estimates ($x_i$*'s*) above as equally accurate ($u_J = u_K = u_L = {}^Gu$ = 0.5%), i.e.**:** $F_J = F_K = F_L = 1$. Then, one can verify**:** $[UF]_C = \Sigma^{L}_{i=J} |M_i^C| = 4.72$, and**:** $[UF]_G = |M_C^G|\,[UF]_C$ = 0.77 (cf. Eq. 7c). That is, the estimate $y_C$ would even then be more **inaccurate** ($\varepsilon_C$ = **4.72**$^Gu$ = 2.36%), but $y_G$ be better **accurate** ($\varepsilon_G$ = **0.77**$^Gu$ = 0.39%), than the measured estimates ($x_i$*'s*). For example**:** $x_J$ = ($X_J$ - 0.5%), $x_K$ = ($X_K$ + 0.5%), and $x_L$ = ($X_L$ - 0.5%), yield "$y_C$ = ($Y_C$ – 2.32%)", which in turn gives "$y_G$ = ($Y_G$ + 0.38%)". However, $x_J$ = ($X_J$ + 0.5%), $x_K$ = ($X_K$ + 0.5%), and $x_L$ = ($X_L$ + 0.5%), yield**:** $y_C = Y_C$, and hence**:** $y_G = Y_G$, which thus confirm the finding[7] that any DTS of the type as Eq. 1b$^{/}$ (here) has got the possibility of leading "$\Delta_i$'s ≠ 0" to "$\delta_d$ = 0".

**3.3 Rate constants ($Y_F$, $Y_D$ and $X_J$) for an enzyme (E) catalyzed reactions of myoglobin (S)**

$$E + S \underset{Y_D}{\overset{Y_F}{\rightleftarrows}} ES \xrightarrow{X_J} E + P$$

The reaction rate constant ($X_J$), and some other parameters ($X_K$, $X_L$, and $X_M$), were obtained[3] by experiments-cum-curve-fitting. However, the formation constant ($Y_F$) and the dissociation constant ($Y_D$) of the ES complex were evaluated[3] as:

$$Y_F = f_F(X_J, X_K, X_L) = \frac{X_K X_L}{X_J C_E^0} \qquad (9)$$

$$Y_D = f_D(X_J, X_K, X_L, Y_F) = X_K + X_L - X_J - Y_F C_E^0 \qquad (9a)$$

where $C_E^0$ is a constant (initial concentration of enzyme). Further, the results of all kinetic parameters were reported[3] to contain ±10% of experimental errors. Therefore, if the uncertainty ($u_J$) in the determined value ($x_J$) of $X_J$ should be 10%, then the uncertainties ($\varepsilon_F$ and $\varepsilon_D$) in the estimates ($y_F$ and $y_D$) of $Y_F$ and $Y_D$ (respectively) should also be 10%. At least, it appears that $\varepsilon_D$, $\varepsilon_F$ and $u_J$, were meant[3] to be equal. However, the idea here is to ascertain the fact.

It may here be mentioned that, though "$X_K$" and "$X_L$" (cf. Eq. 9-9a) were implied[3] to have obtained experimentally, neither their estimates ($x_K$ and $x_L$) nor the corresponding uncertainties ($u_K$ and $u_L$, respectively) were reported. Yet, it may be noted, we first make our consideration (cf. **Block No. 3** in Table 1 for the DTS-specific predicted parameters) for**:** (i) not alone "$X_J$" but $X_K$ and $X_L$ as the measured variables, and (ii) $u_J = u_K = u_L = {}^G u$ (i.e. for**:** $F_J = F_K = F_L = 1$). We, of course, also clarify below the implications for both $X_K$ and $X_L$ to be the constants.

The present system (Eqs. 9-9a) could, if the number $N$ of experimental variables ($X_i$'s) is any yardstick, be considered analogous to the COCP case (Eqs. 8-8a) above. However it should be noted that the features of the 1st stage process as Eq. 9 are predicted to be, unlike those for Eq. 8, independent of $X_i$'s (i.e. $|M_i^F|$ **= 1**, with**:** i = J, K and L, cf. Table 1). Further, that this is the fact can be verified as follows. The actual error ($\delta_F$) in the result ($y_F$) should be obtained as[7]**:** $\delta_F^{Theo} = \sum_{i=J}^{L} M_i^F \Delta_i$ = ($-\Delta_J + \Delta_K + \Delta_L$), i.e. if and when**:** $\Delta_J = (\Delta_K + \Delta_L)$, the result should be absolutely accurate. In fact, it can be seen that a data-set of the type "$x_J = (X_J + 0.05\%)$, $x_K = (X_K + 0.02\%)$, and $x_L = (X_L + 0.03\%)$" but ***irrespective*** of whatever might be the true values of the $X_i$'s yield**:** $y_F = Y_F$ (i.e. error**:** $\delta_F = 0$). These, in addition to confirming that any DTS as Eq. 1b/ can under certain circumstances cause**:** ($\Delta_i$'s $\neq 0$) $\rightarrow$ ($\delta_d = 0$), exemplify the fact[7] that the characteristics of a DTS can (even drastically) vary with its description.

Now, we come to the 2nd stage process (Eq. 9a) but which reduces as:

$$Y_D = X_K + X_L - X_J - (X_K X_L / X_J) = f_D(X_J, X_K, X_L) \neq f_D(Y_F) \qquad (9a^{/})$$

Eq. 9a$^{/}$ clarifies that the dissociation constant ($Y_D$) is governed though differently by the same set of experimental variables ($X_J$, $X_K$, and $X_L$) which define the formation constant ($Y_F$). That is, as**:** $\varepsilon_F = f_F(u_J,u_K,u_L)$, the uncertainty $\varepsilon_D$ can also be independently evaluated as**:** $\varepsilon_D = f_D(u_J,u_K,u_L)$. It may in fact be pointed out that, in Table 1 (Block no. 3), the 2$^{nd}$ stage parameters are furnished against Eq. 9a$^{/}$, and the reason why is that $Y_D$ though which might be computed via Eq. 9a is **not** really a function of $Y_F$. That is to say that, in principle and unlike "$Y_G$" above (cf. Eq. 8a**:** $Y_G = f_G(Y_C) = f_G(X_J,X_K,X_L)$), neither $Y_D$ should be referred as the 2$^{nd}$ stage variable, nor the uncertainty ($\varepsilon_D$) in its estimate ($y_D$) could be obtained as an inter-stage dependent (i.e. COCP) parameter.

However, unlike the case of $Y_F$, the rates ($M_i^D$*'s*, cf. Eq. 2) of variations of $Y_D$ with $X_i$*'s* are predicted to depend on $X_i$*'s* (cf. Table 1). Thus, it may be noted, the tabulated $M_i^D$*-values* were obtained by considering the true $X_i$*'s* to be**:** $X_J = 1.1\ s^{-1}$ (i.e. **as reported**[3]), $X_K = 1.3$, and $X_L = 2.7077\text{x}10^{-8}$ (i.e. **as obtained here** by solving Eq. 9 and Eq. 9a$^{/}$ for[3]**:** $C_E^0 = 3.2\text{x}10^{-6}$ M, $Y_F = 0.01\ M^{-1}\ s^{-1}$, and $Y_D = 0.20\ s^{-1}$). However, though $Y_D$ is practically invariant with $X_L$ ($M_L^D = -2.5\text{x}10^{-8}$, cf. Table 1), the uncertainty in its estimate is predicted to be **4** times and **12** times higher than the uncertainties in the estimates of the forward ($Y_F$) and reaction rate ($X_J$) constants, respectively ($[UF]_D$ **:** $[UF]_F$ **:** $u_J = 12$ **:** 3 **:** 1, cf. Table 1). Thus, for verification, let the evaluation process be so established that**:** $u_i = {}^{G}u = 1\%$. However, though true errors can never be known, imagine that two different sets of experiments have yielded**:** (**1**) $x_J = (X_J - 0.01X_J)$, $x_K = (X_K + 0.01X_K)$, and $x_L = (X_L + 0.01X_L)$; and (**2**) $x_J = (X_J + 0.01X_J)$, $x_K = (X_K - 0.01X_K)$, and $x_L = (X_L - 0.01X_L)$. Then, it could be shown that the set no. **1** gives**:** $y_F = (Y_F + \mathbf{0.0304}Y_F)$, and**:** $y_D = (Y_D + \mathbf{0.12}Y_D)$; and the set no. **2** yields**:** $y_F = (Y_F - \mathbf{0.0296}Y_F)$, and**:** $y_D = (Y_D - \mathbf{0.12}Y_D)$. Further, one can verify that all other possible sets of $x_i$*'s* (with**:** $|{}^{Max}\Delta_i| = \mathbf{1\%}$) imply**:** $|\delta_F| < \mathbf{3\%}$, and**:** $|\delta_D| < \mathbf{12\%}$.

Above, we have considered: $u_J = u_K = u_L = {}^{G}u$. However, if either or both $u_K$ and $u_L$ differ from $u_J$, then the uncertainty factor will also be different. For example, let: $F_J$ = **1**, but: $F_K$ **< 1** and/ or: $F_L$ **< 1**. Then, Eq. 7 predicts: **1 <** $[UF]_F$ **< 3**, and: **5.5 <** $[UF]_D$ **< 12**. Similarly, say "$X_K$" and "$X_L$" are ***constants***, i.e.: $Y_F = f_F(X_J)$; and: $Y_D = f_D(X_J)$. Then the determined[3] $X_J$ and $Y_F$ should certainly be equally accurate ($\varepsilon_F = |M_J^F| u_J = u_J$, cf. Eq. 3a). Yet, as: $|M_J^D| \neq 1$, $\varepsilon_D$ **cannot** be taken to equal $u_J$, i.e. "$y_D$" should be (for the $X_i$ *values* above, 5.5 times more) **inaccurate** than "$x_J$". The behavior of the SSR $Y_F$ (Eq. 9), it may here again be emphasized, is independent of $X_i$*'s*.

### 3.4 Simultaneous determination of carbon and oxygen isotopic abundance ratios as $CO_2^+$

The constituent $^{13}C/^{12}C$, $^{17}O/^{16}O$ and $^{18}O/^{16}O$ abundance ratios (say, $^{III}Y_1$, $^{III}Y_2$ and $^{III}Y_3$, respectively), or even only $^{13}C/^{12}C$ and $^{18}O/^{16}O$ ratios, in $CO_2$ are generally determined via the (inputs $^{II}Y_d$*'s*, and the outputs $Y_d$*'s*, shaping) COCP as below:[10,11]

$$^{I}Y_d = f_d(^{S}X_i, {}^{A}X_i) = (^{S}X_i + 1)\,(^{A}X_i + 1)\,(^{D}K_i + 1) - 1, \quad d = i = 1, 2 = \mathrm{J, K} \qquad (10)$$

$$^{II}Y_d = f_d(^{I}Y_i) = {}^{D}R_i\,(^{I}Y_i + 1), \quad d = i = 1, 2 = \mathrm{J, K} \qquad (11)$$

$$f_i(\{^{III}Y_d\}) = {}^{II}Y_i, \quad (\text{with: } d = 1\text{-}3\text{, but: } i = 1, 2 = \mathrm{J, K}) \qquad (12)$$

$$Y_d = f_d(^{III}Y_d) = (^{III}Y_d / {}^{D}E_d) - 1, \quad d = 1, 3 \text{ (or, } d = 1, 2, 3) \qquad (13)$$

where the prefixes S, A and D refer to the sample $CO_2$, an auxiliary-reference $CO_2$ and a desired-reference $CO_2$, respectively; and $^{D}K_i$, $^{D}R_i$, and $^{D}E_d$, are known constants. Clearly, either Eq. 10 or Eq. 11 or even Eq. 13 represents 2-3 *independent* relationships. However, Eq. 12 is generally referred[10-12] to the following *set* of relationships:

$$^{III}Y_1 + 2\,{}^{III}Y_2 = {}^{II}Y_J \qquad (12a)$$

$$^{III}Y_2\,(2\,{}^{III}Y_1 + {}^{III}Y_2) + 2\,{}^{III}Y_3 = {}^{II}Y_K \qquad (12b)$$

$$^{III}Y_2 = [^{D}E_2 / (^{D}E_3)^{\alpha}]\,(^{III}Y_3)^{\alpha} = \beta(^{III}Y_3)^{\alpha} \qquad (12c)$$

where ${}^{D}E_1$ (cf. Eq. 13), ${}^{D}E_2$ and ${}^{D}E_3$ stand for the known (attributed) values of ${}^{13}C/{}^{12}C$, ${}^{17}O/{}^{16}O$ and ${}^{18}O/{}^{16}O$ abundance ratios, respectively, in the $CO_2$ gas D; $\alpha$ and thus $\beta$ are chosen constants (e.g.**:** $\alpha = 0.5$,[10-12], $\alpha = 0.528$,[13] etc.), so that the solutions for three unknowns (${}^{III}Y_d$*'s*) but from (only two different data on the sample isotopic-molecular abundance ratios ${}^{II}Y_J$ and ${}^{II}Y_K$, and hence from) a set of two equations (nos. 12a-12b) should be possible. Here, it may also be mentioned that the $X_i$-measurement is generally carried out[10,11] as the relative difference between the isotopic molecular abundance ratio (${}^{S}R_i$, or ${}^{A}R_i$) of the $CO_2$ gas (S, or A) of interest and that (${}^{W}R_i$) of a working-reference $CO_2$ gas (W). Thus, here, we refer (cf. Eq. 10)**:**[10]

$${}^{S}X_J = [({}^{S}R_J / {}^{W}R_J) - 1] = [({}^{S}R_{45/44} / {}^{W}R_{45/44}) - 1];\ \text{and}\ {}^{A}X_J = [({}^{W}R_{45/44} / {}^{A}R_{45/44}) - 1];$$

$${}^{S}X_K = [({}^{S}R_{46/44} / {}^{W}R_{46/44}) - 1];\ \text{and}\ \ {}^{A}X_K = [({}^{W}R_{46/44} / {}^{A}R_{46/44}) - 1]).$$

Similarly**:** ${}^{D}K_i = [({}^{A}R_i / {}^{D}R_i) - 1]$. Further, even for given ${}^{D}E_2$ and ${}^{D}E_3$ (cf. Eq. 12c), $\alpha$ will vary with the sample (i.e. ${}^{III}Y_2$ and ${}^{III}Y_3$). That is to say that there can be no way to exactly preset "$\alpha$". In other words, not alone the measured estimates (${}^{S}x_J$, ${}^{A}x_J$, ${}^{S}x_K$, and ${}^{A}x_K$) should be at certain uncertainties (${}^{S}u_J$, ${}^{A}u_J$, ${}^{S}u_K$, and ${}^{A}u_K$, respectively) but even the chosen value of $\alpha$ would be subject to some uncertainty $u_\alpha$. We therefore treat $\alpha$ as a measured cum input variable ($i = L$), thereby offering a means to study the effects of its variation on the desired results.

The COCP-parameters ($M_i^d$*'s*, $[UF]_d$*'s,* and $\varepsilon_d$*'s*) are furnished in Table 2. The case of the somewhat involved process as Eq. 12 is even elaborated in APPENDIX 1. However, as clarified (cf. Table 2 and Appendix 1), the characteristics of all the SSRs (Eqs. 10-13) will vary with the measured variables and the constants as well. Thus, it may be noted that the numerical values of $M_i^d$*'s* in Table 2 are referred to[10]**:** ${}^{S}X_J = -0.020$, ${}^{A}X_J = -0.0070$, ${}^{D}K_J = 1.756367272\text{x}10^{-3}$, and ${}^{S}X_K = 0.020$, ${}^{A}X_K = -0.0010$, ${}^{D}K_K = -2.193768974\text{x}10^{-3}$ **(cf. 1st stage)**, ${}^{D}R_J = 11.99880669\text{x}10^{-3}$, and ${}^{D}R_K = 41.85401492\text{x}10^{-4}$ **(cf. 2nd stage)**, and ${}^{D}E_1 = 11.2372\text{x}10^{-3}$, ${}^{D}E_2 = 38.0803342\text{x}10^{-5}$, ${}^{D}E_3 =$

20.88349077x$10^{-4}$, and $\alpha$ = 0.5 (cf. **3rd** and **4th stages)**. In addition, the $[UF]_d$, and hence the $\varepsilon_d$, values are predicted for: $^{S}u_J = {}^{A}u_J = {}^{S}u_K = {}^{A}u_K = u_\alpha = {}^{G}u$, i.e. for: $F_i(s)$ = **1.0**. Therefore, "$^{G}u$" in Table 2 could be read as "$u_i$ (i = J, K, or $\alpha$)", e.g.: $^{I}\varepsilon_J = 1.06u_i$. Further, if: $^{G}u$ = 1%, then, for a specified $Y_d$, "$\varepsilon_d$" will numerically equal to "$[UF]_d$". Such a fact is exemplified in Table 3, which furnishes, for: $\boldsymbol{x_i} = (\boldsymbol{X_i} \pm \mathbf{0.01}\boldsymbol{X_i})$, all the different stage-outputs ($y_d$'s) and their actual errors ($\delta_d$'s). It may be noted that only two specific examples (nos. **1** and **2**) yield, like the cases discussed above: $|\delta_d| = \varepsilon_d = [UF]_d$, but corresponding others (nos. **3-5**) give: $|\delta_d| \leq \varepsilon_d$.

What may however be worth pointing out is that, even for a given input-uncertainty ($u_i = {}^{G}u$, with: $i = J$, $K$ and, $\alpha$), $\varepsilon_d \neq {}^{III}\varepsilon_d \neq {}^{II}\varepsilon_d \neq {}^{I}\varepsilon_d \neq u_i$ (cf. Table 2), which supplement the previous finding[7] that no DTS (Eq. 1 or COCP) can be without checking its properties taken to yield the desired output(s) exactly so accurate as its input(s). In other words, the study here clarifies that the comparison between labs on the measured data ($X_i$-*values*), and that on the corresponding results (scale-converted data: $Y_d$-*values*), may not stand the same. For illustration, suppose that two different labs: (i) are equipped for measurements with **1%** accuracy ($^{G}u = u_i$ = 1%), and (ii) have reported the measured data, and hence the results, on their independently collected samples, as those described by the example nos. 1 and 2, respectively, in Table 3. Then, it may be noted that the measured data, e.g. $^{S}x_J = ({}^{S}X_J + 0.01X_J) = -0.0202$ and, $^{S}x_J = ({}^{S}X_J - 0.01X_J) = -0.0198$ (cf. example nos. 1 and 2, respectively), imply a variation of **2%** between labs. However, it may be pointed out, the lab to lab output-variation depends on the *stage no.* and the output-*variable* as well, e.g. the COCP results as $y_1$ differ by **2.18%**, but $y_2$-values by **4.5%** and the $y_3$-values by **2.5**%. On the contrary, the variation between the outputs for a specified 2nd or 3rd stage variable appeals relatively insignificant. For example, the $^{III}y_2$ values vary from one another by the least extent (**0.038%**), even though the desired $y_2$-values are most varied (**4.5%**), among all.

Now, it may also be noted, the example no. 6 (with**:** $\alpha = 0.53$) in Table 3 is different from the example no. 0 (with**:** $\alpha = 0.50$) only for $\alpha$. However, the variation is significant ($\Delta_\alpha = 6\%$). Yet, it may be pointed out, the corresponding estimates (i.e. ${}^{III}y_d$*'s*) are at relatively insignificant errors. This verifies the prediction in Table 2 (cf. the ${}^{III}M_\alpha^d$-values) and/ or the previous report[14] that ${}^{III}Y_d$*'s* are much less sensitive towards the variation in $\alpha$ than that in the measured data.

## 4. CONCLUSIONS

It is above shown how, in the case of an indirect measurement system represented specifically by a COCP (viz.**:** $X_J \rightarrow {}^{I}Y_d \xrightarrow{X_K} {}^{II}Y_d \ldots \xrightarrow{X_N} Y_d$), the uncertainty ($\varepsilon_d$) in any stage-specific or COCP estimate ($y_d$), and hence the requirements for $y_d$ to be as representative as desired, can a priori be predicted. Basically, the study only helps generalize a previous finding[7] that the transformation of the measured estimate(s) $x_i(s)$ into the desired estimate(s) $y_d(s)$ **(**using the given (single set of, or cascade of) system-specific relationship(s) of the measured $X_i(s)$ with the desired $Y_d(s)$**)** can cause the $y_d(s)$ to worse or even better represent the evaluation-system than the $x_i(s)$. It is thus exemplified above that, while the 1st stage result (${}^{I}y_d$) could depending upon the nature of the corresponding SSR be more inaccurate than the measured data (${}^{I}\varepsilon_d > u_i$), the 2nd stage result might turn out more accurate (${}^{II}\varepsilon_d < u_i$). In other words, it is for alone the characteristic behavior(s) of the SSR(s) and/ or COCP representing the evaluation-system shown to be possible that the measured data and the corresponding derived (scale-converted) output(s) lead to quite diverging conclusions about the system/ sample studied. The a priori uncertainty ($\varepsilon_d$) evaluation shown above can however help to avoid such a possible confusion, and even properly preplan the required experiments, thereby arriving at the truth sought.

Further, the theoretical tool as an SSR is clarified to behave like a physical measuring device (MD), and hence is also referred to above as a data transformation scale (DTS). However, while a given evaluation refers a given (set of) DTS(s), there could always be more than one MD enabling the required $X_i$-measurement(s). However the one, ensuring the achievable accuracy to be not alone high (i.e. offering $u_i$ to be close to zero) but unvarying as a function of $X_i$, should generally be the preferred technique. That is to say that the *uniformity in behavior* should be a desired characteristic for an MD to be chosen in practice. However, even for given measurement-accuracy(s) $u_i(s)$, it clarified above that accuracy $\varepsilon_d$ can for one or the other given case of DTS considerably vary with *alone* the $X_i$*-value(s). Non-uniformity in behavior* is thus attributed to be the characteristic for the DTS in general, thereby emphasizing the need for a priori establishment of not simply the method-specific $u_i$ but the nature of the DTS (i.e. $M_i^d$-values, and hence $\varepsilon_d$).

**APPENDIX 1: $^{III}M_i^d$'s and $^{III}[UF]_d$'s corresponding to the set of DTS as Eqs. 12a-12c**

Eqs. 12a-12c can in differential form be expressed as Eqs. A.1-A.3, respectively:[14]

$$C_{J1}\,\partial\,^{III}Y_1 + C_{J2}\,\partial\,^{III}Y_2 = C_J\,\partial\,^{II}Y_J \tag{A.1}$$

$$C_{K1}\,\partial\,^{III}Y_1 + C_{K2}\,\partial\,^{III}Y_2 + C_{K3}\,\partial\,^{III}Y_3 = C_K\,\partial\,^{II}Y_K \tag{A.2}$$

$$C_{\alpha 2}\,\partial\,^{III}Y_2 + C_{\alpha 3}\,\partial\,^{III}Y_3 = C_\alpha\,\partial\alpha \tag{A.3}$$

where: $C_{J1} = 1$, $C_{J2} = 2$, and $C_J = 1$ (cf. **Eq.12a**); $C_{K1} = 2\,^{III}Y_2$, $C_{K2} = 2(^{III}Y_1 + {}^{III}Y_2)$, $C_{K3} = 2$, and $C_K = 1$ (cf. **Eq. 12b**); and: $C_{\alpha 2} = (1 / {}^{III}Y_2)$, $C_{\alpha 3} = (-\alpha / {}^{III}Y_3)$, and $C_\alpha = \ln(^{III}Y_3 / {}^{D}E_3)$, (cf. **Eq.12c**).

Now let: $Q = [C_{J1}\,(C_{K2}\,C_{\alpha 3} - C_{K3}\,C_{\alpha 2}) - C_{J2}\,C_{K1}\,C_{\alpha 3}]$; $N_{JJ} = (^{II}Y_J\,C_J / Q)$; $N_{KK} = (^{II}Y_K\,C_K / Q)$; and: $N_{\alpha\alpha} = (\alpha\,C_\alpha / Q)$, then it could be shown that:

$$^{III}M_J^1 = [(C_{K2}\,C_{\alpha 3} - C_{K3}\,C_{\alpha 2})\,N_{JJ} / {}^{III}Y_1] \tag{A.4}$$

$$^{III}M_K^1 = (-\,C_{J2}\,C_{\alpha 3}\,N_{KK} / {}^{III}Y_1) \tag{A.4a}$$

$$^{III}M_\alpha^1 = (C_{J2}\,C_{K3}\,N_{\alpha\alpha} / {}^{III}Y_1) \tag{A.4b}$$

$$^{III}M_J^2 = (-\,C_{K1}\,C_{\alpha 3}\,N_{JJ} / {}^{III}Y_2) \tag{A.5}$$

$$^{III}M_K^2 = (C_{J1}\,C_{\alpha 3}\,N_{KK} / {}^{III}Y_2) \tag{A.5a}$$

$$^{III}M_\alpha^2 = (-\,C_{J1}\,C_{K3}\,N_{\alpha\alpha} / {}^{III}Y_2) \tag{A.5b}$$

$$^{III}M_1^3 = (C_{K1}\,C_{\alpha 2}\,N_{JJ} / {}^{III}Y_3) \tag{A.6}$$

$$^{III}M_2^3 = (-\,C_{J1}\,C_{\alpha 2}\,N_{KK} / {}^{III}Y_3) \tag{A.6a}$$

$$^{III}M_\alpha^3 = [(C_{J1}\,C_{K2} - C_{J2}\,C_{K1})\,N_{\alpha\alpha} / {}^{III}Y_3] \tag{A.6b}$$

Further:

$^{III}[UF]_1 = (\,|\,^{III}M_J^1\,|\,^{II}[UF]_J + |\,^{III}M_K^1\,|\,^{II}[UF]_K + |\,^{III}M_\alpha^1\,|\,F_\alpha)$, with: $F_\alpha = (u_\alpha / {}^{G}u)$;

$^{III}[UF]_2 = (\,|\,^{III}M_J^2\,|\,^{II}[UF]_J + |\,^{III}M_K^2\,|\,^{II}[UF]_K + |\,^{III}M_\alpha^2\,|\,F_\alpha)$; and

$^{III}[UF]_3 = (\,|\,^{III}M_J^3\,|\,^{II}[UF]_J + |\,^{III}M_K^3\,|\,^{II}[UF]_K + |\,^{III}M_\alpha^3\,|\,F_\alpha)$.

**Table 1:** Stage (DTS) specific parameters for three different COCP systems: Eqs. (5-5b), Eqs. (8-8a), and Eqs (9-9a)

| Block No. | Stage No. (DTS) | $Y_d$ | $M_i^d$ (cf. Eq. 2) | $[UF]_d$ (cf. Eq. 7/ 7a/ 7b) | $\varepsilon_d$ (cf. Eq. 6/ 6a/ 6b) |
|---|---|---|---|---|---|
| 1 | I (Eq. 5) | ${}^{I}Y_d$ = 2.0 | ${}^{I}M_J^d = X_J / (X_J - X_K) = 2.5$, and: ${}^{I}M_K^d = -X_K / (X_J - X_K) = -1.5$ | ${}^{I}[UF]_d = \Sigma^{K}_{i=J} \lvert {}^{I}M_i^d \rvert F_i = \Sigma^{K}_{i=J} \lvert {}^{I}M_i^d \rvert = \mathbf{4}$ | ${}^{I}\varepsilon_d = {}^{I}[UF]_d\ {}^{G}u = 4\ {}^{G}u = \mathbf{0.2}\%$ |
| | II (Eq. 5a) | ${}^{II}Y_d$ = 0.75 | ${}^{II}M_d^d = {}^{I}Y_d / ({}^{I}Y_d + 1) = 0.6667$ | ${}^{II}[UF]_d = \lvert {}^{II}M_d^d \rvert\ {}^{I}[UF]_d = 0.6667 \times 4 = \mathbf{2.667}$ | ${}^{II}\varepsilon_d = {}^{II}[UF]_d\ {}^{G}u = 2.667\ {}^{G}u = \mathbf{0.133}\%$ |
| | III (Eq. 5b) | ${}^{III}Y_d$ = 1.25 | ${}^{III}M_d^d = {}^{II}Y_A / ({}^{II}Y_d + X_L) = 0.6$, and: ${}^{III}M_L^d = X_L / ({}^{II}Y_d + X_L) = 0.4$ | ${}^{III}[UF]_d = \lvert {}^{III}M_d^d \rvert\ {}^{II}[UF]_d + \lvert {}^{III}M_L^d \rvert F_L = (0.6 \times 2.667) + (0.4 \text{ x } 8) = \mathbf{4.8}$ | ${}^{III}\varepsilon_d = {}^{III}[UF]_d\ {}^{G}u = 4.8\ {}^{G}u = \mathbf{0.24}\%$ |
| 2 | I (Eq. 8) | $Y_C$ = 28.2 mmol dm$^{-3}$ | $M_J^C = 1.0$, $M_K^C = -X_K / (X_K - X_L) = -2.36$, and: $M_L^C = X_L / (X_K - X_L) = 1.36$ | $[UF]_C = \Sigma^{L}_{i=J} \lvert M_i^C \rvert F_i = \mathbf{54.4}$ | $\varepsilon_C = [UF]_C\ {}^{G}u = 54.44\ {}^{G}u = \mathbf{27.2}\%$. |
| | II (Eq. 8a) | $Y_G$ = -15.1 KJ mol$^{-1}$ | $M_C^G = (\boldsymbol{A}\ ln\ (\boldsymbol{B}/\boldsymbol{A}))^{-1} = -0.1637$ (with: $\boldsymbol{A} = [1 + V_0\, Y_C]$, and: $\boldsymbol{B} = V_0\, Y_C$) | $[UF]_G = \lvert M_C^G \rvert\ [UF]_C = \mathbf{8.9}$ | $\varepsilon_G = [UF]_G\ {}^{G}u = 8.9\ {}^{G}u = \mathbf{4.45}\%$ |
| 3 | I (Eq. 9) | $Y_F$ = 0.01 | $M_J^F = \mathbf{-1}$, $M_K^F = \mathbf{1}$, and: $M_L^F = \mathbf{1}$ | $[UF]_F = \Sigma^{L}_{i=J} \lvert M_i^F \rvert F_i = \Sigma^{L}_{i=J} \lvert M_i^F \rvert = \mathbf{3}$ | $\varepsilon_F = [UF]_F\ {}^{G}u = \mathbf{3}\,{}^{G}u$ |
| | II (Eq. 9a$^{/}$) | $Y_D$ = 0.2 | $M_J^D = (X_K X_L - X_J^2) / (X_J Y_D) = -5.5$, $M_K^D = X_K (X_J - X_L) / (X_J Y_D) = 6.5$, and: $M_L^D = X_L (X_J - X_K) / (X_J Y_D) = -2.5\text{x}10^{-8}$ | $[UF]_D = \Sigma^{L}_{i=J} \lvert M_i^D \rvert F_i = \Sigma^{L}_{i=J} \lvert M_i^D \rvert = \mathbf{12}$ | $\varepsilon_D = [UF]_D\ {}^{G}u = \mathbf{12}\,{}^{G}u$ |

**Table 2:** Stage (DTS) specific parameters for the (COCP as Eqs. 10-13**:)** determination of constituent carbon and oxygen isotopic abundance ratios in the sample $CO_2$ gas (S)

| Stage No. (DTS) | $Y_d$ | $M_i^d$ (cf. Eq. 2) | $[UF]_d$ (cf. Eq. 6/ 6a/ 6b) | $\varepsilon_d$ (cf. Eq. 5/ 5a/ 5b) |
|---|---|---|---|---|
| I (Eq. 10) | ${}^{I}Y_J$ | ${}^{S}M_J^J = {}^{S}X_J\,({}^{A}X_J + 1)\,({}^{D}K_J + 1) / {}^{I}Y_J = 0.791$<br>${}^{A}M_J^J = {}^{A}X_J\,({}^{S}X_J + 1)\,({}^{D}K_J + 1) / {}^{I}Y_J = 0.273$ | ${}^{I}[UF]_J = \lvert {}^{S}M_J^J \rvert\, {}^{S}F_J + \lvert {}^{A}M_J^J \rvert\, {}^{A}F_J = \lvert {}^{S}M_J^J \rvert + \lvert {}^{A}M_J^J \rvert = 1.06$ | ${}^{I}\varepsilon_J = {}^{I}[UF]_J\, {}^{G}u = 1.06\, {}^{G}u$ |
| | ${}^{I}Y_K$ | ${}^{S}M_K^K = {}^{S}X_K\,({}^{A}X_K + 1)\,({}^{D}K_K + 1) / {}^{I}Y_K = 1.19$<br>${}^{A}M_K^K = {}^{A}X_K\,({}^{S}X_K + 1)\,({}^{D}K_K + 1) / {}^{I}Y_K = -0.0608$ | ${}^{I}[UF]_K = \lvert {}^{S}M_K^K \rvert\, {}^{S}F_K + \lvert {}^{A}M_K^K \rvert\, {}^{A}F_J = \lvert {}^{S}M_K^K \rvert + \lvert {}^{A}M_K^K \rvert = 1.25$ | ${}^{I}\varepsilon_K = {}^{I}[UF]_K\, {}^{G}u = 1.25\, {}^{G}u$ |
| II (Eq. 11) | ${}^{II}Y_J$ | ${}^{II}M_J^J = {}^{I}Y_J / ({}^{I}Y_J + 1) = -0.0258$ | ${}^{II}[UF]_J = \lvert {}^{II}M_J^J \rvert\, {}^{I}[UF]_J = 0.027$ | ${}^{II}\varepsilon_J = {}^{II}[UF]_J\, {}^{G}u = 0.027\, {}^{G}u$ |
| | ${}^{II}Y_K$ | ${}^{II}M_K^K = {}^{I}Y_K / ({}^{I}Y_K + 1) = 0.0165$ | ${}^{II}[UF]_K = \lvert {}^{II}M_K^K \rvert\, {}^{I}[UF]_K = 0.021$ | ${}^{II}\varepsilon_K = {}^{II}[UF]_K\, {}^{G}u = 0.021\, {}^{G}u$ |
| III [*1] (Eqs. 12a-12c) | ${}^{III}Y_1$ | ${}^{III}M_J^1 = 1.07$, ${}^{III}M_K^1 = -0.0352$, and ${}^{III}M_\alpha^1 = -5.85\text{x}10^{-4}$ | ${}^{III}[UF]_1 = 0.031$ | ${}^{III}\varepsilon_1 = 0.031\, {}^{G}u$ |
| | ${}^{III}Y_2$ | ${}^{III}M_J^2 = -0.0011$, ${}^{III}M_K^2 = 0.501$, and ${}^{III}M_\alpha^2 = 0.0083$ | ${}^{III}[UF]_2 = 0.019$ | ${}^{III}\varepsilon_2 = 0.019\, {}^{G}u$ |
| | ${}^{III}Y_3$ | ${}^{III}M_J^3 = -0.0021$, ${}^{III}M_K^3 = 1.001$, and ${}^{III}M_\alpha^3 = -1.6\text{x}10^{-5}$ | ${}^{III}[UF]_3 = 0.021$ | ${}^{III}\varepsilon_3 = 0.021\, {}^{G}u$ |
| IV (Eq. 13) | $Y_1$ | $M_1^1 = {}^{III}Y_1 / ({}^{III}Y_1 - {}^{D}E_1) = -35.5$ | $[UF]_1 = \lvert M_1^1 \rvert\, {}^{III}[UF]_1 = 1.09$ | $\varepsilon_1 = [UF]_1\, {}^{G}u = 1.09\, {}^{G}u$ |
| | $Y_2$ | $M_2^2 = {}^{III}Y_2 / ({}^{III}Y_2 - {}^{D}E_2) = 120.4$ | $[UF]_2 = \lvert M_2^2 \rvert\, {}^{III}[UF]_2 = 2.25$ | $\varepsilon_2 = [UF]_2\, {}^{G}u = 2.25\, {}^{G}u$ |
| | $Y_3$ | $M_3^3 = {}^{III}Y_3 / ({}^{III}Y_3 - {}^{D}E_3) = 60.5$ | $[UF]_3 = \lvert M_3^3 \rvert\, {}^{III}[UF]_3 = 1.25$ | $\varepsilon_3 = [UF]_3\, {}^{G}u = 1.25\, {}^{G}u$ |

**[*1]:** See APPENDIX 1

**Table 3:** Examples of variations in the stage specific outputs (for simultaneously determining carbon and oxygen isotopic abundance ratios as $CO_2^+$) as a function of variations in the inputs ($^{S}X_J$, $^{A}X_J$, $^{S}X_K$, $^{A}X_K$ and, $\alpha$)

| Example No. | $^{I}y_J \times 10^3$ (% $^{I}\delta_J$) | $^{I}y_K \times 10^3$ (% $^{I}\delta_K$) | $^{II}y_J \times 10^3$ (% $^{II}\delta_J$) | $^{II}y_K \times 10^4$ (% $^{II}\delta_K$) | $^{III}y_1 \times 10^3$ (% $^{III}\delta_1$) | $^{III}y_2 \times 10^5$ (% $^{III}\delta_2$) | $^{III}y_3 \times 10^4$ (% $^{III}\delta_3$) | $y_1 \times 10^3$ (% $\delta_1$) | $y_2 \times 10^4$ (% $\delta_2$) | $y_3 \times 10^3$ (% $\delta_3$) |
|---|---|---|---|---|---|---|---|---|---|---|
| 0 [*0] | -25.1508 (0.0) | 16.7446 (0.0) | | | | | | -27.4230 (0.0) | | 16.8184 (0.0) |
| 00 [*00] | -25.1508 (0.0) | 16.7446 (0.0) | 11.6970 (0.0) | 42.5548 (0.0) | 10.929043 (0.0) | 38.399224 (0.0) | 21.234718 (0.0) | -27.422978 (0.0) | 83.741391 (0.0) | 16.818404 (0.0) |
| 1 [*1] | -25.4185 (1.06) | 16.9541 (1.25) | 11.6938 (-0.027) | 42.5636 (0.021) | 10.925687 (-0.031) | 38.406435 (0.019) | 21.239108 (0.021) | -27.721608 (1.09) | 85.635087 (2.26) | 17.028609 (1.25) |
| 2 [*2] | -24.8831 (-1.06) | 16.5351 (-1.25) | 11.7002 (0.027) | 42.5461 (-0.021) | 10.932397 (0.031) | 38.392093 (-0.019) | 21.230328 (-0.021) | -27.124460 (-1.09) | 81.868677 (-2.24) | 16.608200 (-1.25) |
| 3 [*3] | -25.4185 (1.06) | 16.9541 (1.25) | 11.6938 (-0.027) | 42.5636 (0.021) | 10.925816 (-0.030) | 38.399957 (0.0019) | 21.239114 (0.021) | -27.710078 (1.05) | 83.933852 (0.23) | 17.028936 (1.25) |
| 4 [*4] | -25.2811 (0.52) | 16.5351 (-1.25) | 11.6955 (-0.013) | 42.5461 (-0.021) | 10.927622 (-0.013) | 38.392109 (-0.019) | 21.230346 (-0.021) | -27.549384 (0.46) | 81.872982 (-2.23) | 16.609077 (-1.24) |
| 5 [*5] | -25.4185 (1.06) | 16.9338 (1.13) | 11.6938 (-0.027) | 42.5628 (0.019) | 10.925695 (-0.031) | 38.406047 (0.018) | 21.238682 (0.019) | -27.720916 (1.09) | 85.533014 (2.14) | 17.008227 (1.13) |
| 6 [*6] | -25.1508 (0.0) | 16.7446 (0.0) | 11.6970 (0.0) | 42.5548 (0.0) | 10.928659 (-0.0035) | 38.418423 (**0.05**) | 21.234698 (-0.000095) | -27.457148 (0.125) | 88.783019 (**6.02**) | 16.817435 (-0.0058) |

[*0]**:** Results as reported[10] for**:** $^{S}X_J = -0.020$, $^{A}X_J = -0.0070$, $^{S}X_K = 0.020$, $^{A}X_K = -0.0010$, and $\alpha = 0.5$ (cf. the text).

[*00]**:** A verification that our evaluated results (for**:** $\boldsymbol{x_i(s) = X_i(s)}$, i.e. for**:** $\boldsymbol{\Delta_i(s) = 0.0}$) are the same as given in [10].

[*1]**:** $^{S}x_J = -0.0202$, $^{A}x_J = -0.00707$, $^{S}x_K = 0.0202$, $^{A}x_K = -0.00099$, and $\alpha = 0.5050$ (i.e.**:** $^{S}\Delta_J = {}^{A}\Delta_J = {}^{S}\Delta_K = \Delta_\alpha = 1\%$ and**:** $^{A}\Delta_K = -1\%$).

[*2]**:** Results for "$^{S}\Delta_J = {}^{A}\Delta_J = {}^{S}\Delta_K = \Delta_\alpha = -1\%$ and**:** $^{A}\Delta_K = 1\%$".

[*3]**:** $^{S}\Delta_J = {}^{A}\Delta_J = {}^{S}\Delta_K = 1\%$, and**:** $^{A}\Delta_K = \Delta_\alpha = -1\%$.

[*4]**:** $^{S}\Delta_J = {}^{A}\Delta_K = 1\%$, and**:** $^{S}\Delta_K = {}^{A}\Delta_J = \Delta_\alpha = -1\%$.

[*5]**:** $^{S}\Delta_J = {}^{A}\Delta_J = {}^{S}\Delta_K = {}^{A}\Delta_K = \Delta_\alpha = 1\%$.

[*6]**:** $^{S}\Delta_J = {}^{A}\Delta_J = {}^{S}\Delta_K = {}^{A}\Delta_K = \mathbf{0.0}$, but $\Delta_\alpha = \mathbf{6\%}$.